\documentclass[letterpaper, 10 pt, conference]{ieeeconf}  % Comment this line out
\IEEEoverridecommandlockouts                              % This command is only
\usepackage{graphicx} % for pdf, bitmapped graphics files

\title{\LARGE \bf Achieving Data Dissemination with Security using FIWARE and Intel Software Guard Extensions (SGX)}

\author{Dalton C\'ezane Gomes Valadares$^{1,2}$, Matteus Sthefano Leite da Silva$^{2}$,\\ Andrey El\'isio Monteiro Brito$^{2}$ and Ewerton Monteiro Salvador$^{3}$
\thanks{*This research was partially funded by EU-BRA SecureCloud project
(EC, MCTIC/RNP, and SERI, 3rd Coordinated Call, H2020-ICT-2015
Grant agreement no. 690111)}%
\thanks{$^{1}$Federal University of Campina Grande, Informatics and Electrical Engineering Center, Computer Science, Campina Grande, Paraíba, Brazil
        {\tt\small cezane@lsd.ufcg.edu.br, silvamatteus@lsd.ufcg.edu.br, andrey@computacao.ufcg.edu.br}}
\thanks{$^{2}$Federal Institute of Pernambuco, Mechanical Engineering Department, Caruaru, Pernambuco, Brazil
        {\tt\small dalton.valadares@caruaru.ifpe.edu.br}},
\thanks{$^{3}$Federal University of Para\'iba, Jo\~ao Pessoa, Para\'iba, Brazil
        {\tt\small esalvador@lsd.ufcg.edu.br}}
\thanks{\copyright \copyright 2018 IEEE. Personal use of this material is permitted. Permission from IEEE
must be obtained for all other uses, in any current or future media, including
reprinting/republishing this material for advertising or promotional purposes,
creating new collective works, for resale or redistribution to servers or lists, or reuse
of any copyrighted component of this work in other works.}
}

\begin{document}

\maketitle
\thispagestyle{empty}
\pagestyle{empty}

%%%%%%%%%%%%%%%%%%%%%%%%%%%%%%%%%%%%%%%%%%%%%%%%%%%%%%%%%%%%%%%%%%%%%%%%%%%%%%%%
\begin{abstract}

The Internet of Things (IoT) field has gained much attention from industry and academia, being the main subject for numerous research and development projects. Frequently, the dense amount of generated data from IoT applications is sent to a cloud service, that is responsible for processing and storage. Many of these applications demand security and privacy for their data because of their sensitive nature. This is specially true when such data must be processed in entities hosted in public clouds, where the environment in which applications run may not be trusted. Some concerns are then raised since it is not trivial to provide the needed protection for these sensitive data. We present a solution that considers the security components of FIWARE and the Intel SGX capabilities. FIWARE is a platform created to support the development of Smart Applications, including IoT systems, and SGX is the Intel solution for Trusted Execution Environment (TEE). We propose a new component for key management that, together with other FIWARE components, can be used to provide privacy, confidentiality, and integrity guarantees for IoT data. A case study illustrates how this proposed solution can be employed in a realistic scenario, which allows the dissemination of sensitive data through public clouds without risking privacy issues. The results of the experiments provide evidence  that our approach does not harm scalability or availability of the system. In addition, it presents acceptable memory costs when considering the benefit of the privacy guarantees achieved.

\end{abstract}

%%%%%%%%%%%%%%%%%%%%%%%%%%%%%%%%%%%%%%%%%%%%%%%%%%%%%%%%%%%%%%%%%%%%%%%%%%%%%%%%
\section{Introduction}
The Internet of Things (IoT) concept has gained considerable importance in industry and academia in the last years. The possibility of interconnecting a huge number of devices allows the creation of various new applications. This is possible thanks to the advances in devices technologies and communication protocols that enable such interconnection. IoT sensors and actuators are increasingly becoming smarter and, even when they are not so smart, they can connect to other gateway devices that may add additional functionality.

IoT devices often have some common constraints like limited storage and processing capacities, limited energy, or connectivity. These constraints are dealt by cloud-based IoT. The idea, in general, is to send all collected data to the cloud in a way they can be analyzed and processed for use in other systems~\cite{Li2016}. Among the cloud-based IoT benefits, the following can be highlighted: long-term storage and processing of collected data, data reuse in multiple services, data integration of different users/things, user mobility support and granted availability for intermittent connection~\cite{Henze:2016:CAP:2869179.2869387}.

Frequently, the IoT applications will be generating sensitive data related to specific businesses that can involve large financial transactions, or even related to personally identifiable information (PII) of their users. Both cases, among many others, require special attention in order to provide security and privacy capabilities for these kind of applications. It is mandatory to ensure an acceptable level of trust for processing and storing these data and this is an actual challenge nowadays (privacy and security are still concerns related to IoT applications \cite{Li2016,wilson2017}).

A common scenario for an IoT application considers a publish/subscribe system that is responsible for registering the interest of some entities in specific data or measurements produced by another and for sending notifications to the interested entities whenever changes occur in the measured values. When the notifications are received, the entities can take important decisions as turning on/off machines, triggering alarms, performing complex processing, etc. 

An instance of this previous scenario can be an energy consumption monitoring application, considering two perspectives to facilitate the understanding: firstly, there are many smart meters, named here \textit{Producers}, collecting energy consumption from different buildings along a smart city; secondly, there are applications responsible for doing the data aggregation for customers' billing purposes. Some techniques that hit the customers' privacy, like NIALM \cite{Eibl2014}, enable attackers to estimate the number of people in buildings, as well as to gain information on what kinds of activities are being performed. The same can be applied to an industrial environment, in which attackers can discover secrets related to production, estimating how many machines are working, how many workers, etc. Because of this, it is highly recommended to keep a good level of security and privacy in architectures that support such scenarios.

To avoid such kind of attacks, the data can be encrypted in the producers in order to be processed only in a secure way. This can be achieved by homomorphic cryptography~\cite{Silva2017}, which performs common operations over encrypted data, but its great overhead makes it unpractical to use with complex operations~\cite{SCBR}.
%%%% TODO: fazer uma referência a SCBR não parece adequado aqui, não é melhor fazer a Leandro de novo? estava SCBR
Another approach is the use of a trusted execution environment (TEE), that enables, by using specific hardware instructions, the creation of a shielded space in the memory in a way data can be processed securely inside this space. Encrypted data enters the protected area, being decrypted only inside it and encrypted again before leaving. TEE use is more feasible than homomorphic cryptography, because, it takes considerably less time to process data \cite{Silva2017}.

Related to this problem, two research questions were identified for guiding this research:
\begin{enumerate}
\item How to limit the need to trust the storage provider and the consumer of the data? 
\item How to control access allowing only authorized entities to consume data?
\end{enumerate}

In this paper we present a solution that prevent users' data from unauthorized access (e.g., NIALM attacks), keeping them private and secure. For this, we considered both previous questions. First, we use the Intel TEE technology, named Software Guard Extensions (SGX). We propose a component responsible for key generation, storage and management, named Key Vault. Key Vault also provides data security by using Intel SGX to protect its execution. The producers send the encrypted measurements for a publish/subscribe component and the consumers, that are also SGX applications, receive these measurements through notifications. Thus, the measurements are protected because they are only decrypted and processed in a secure space, inside an SGX application. 

Second, we have used some components from FIWARE, which is a software framework that provides modules to ease the development of smart applications. For authentication and authorization of producers and consumers, we have used the components Keyrock Identity Management and Wilma PEP Proxy. For the publish/subscribe system we have used the FIWARE Orion Context Broker, which receives the encrypted measurements from producers and triggers notifications to consumers.

This paper is organized as follows: in Section~\ref{sec:background} we present a background, concisely explaining FIWARE, Intel SGX and our case study; the proposed solution is presented in Section~\ref{sec:solution} including principles, architecture, threat model and the definition of Key Vault; the evaluation of our solution as well as the results and a discussion of them are presented in Section~\ref{sec:eval}; in Section~\ref{sec:rwork}, we discuss the related works and finish this paper with Section~\ref{sec:conclusion}, presenting the conclusions and suggesting future works.
\section{Background}
\label{sec:background}
 
In this Section we present some basic concepts in order to ease the understanding of our proposal. FIWARE and Intel Software Guard Extensions (SGX) are shortly covered next. In addition, our case study is briefly described.
%with its scenario and its problems.

\subsection{FIWARE}

FIWARE (Future Internet-WARE) is a software platform created to ease the development of smart applications for the Future Internet in multiple sectors and following open standards~\cite{FIWARE}. For this, it provides a wide set of APIs (Application Programming Interfaces).
% that simple to use in a standardized way. 
FIWARE is composed by many components, called Generic Enablers (GEs), with some of them being based in the well-accepted cloud platform OpenStack. OpenStack is also used to establish a cloud that provides FIWARE services. For each FIWARE component (GE) specification, there is an open source reference implementation. Some of the domains covered by FIWARE GEs are the following~\cite{FIWARECat}: Data and Context Management, Internet of Things Services Enablement, Security, Cloud Hosting, Advanced Web-based User Interface, etc.

This work uses three FIWARE GEs for our proof of concept: Orion Context Broker (CB), Keyrock Identity Manager and Wilma PEP (Policy Enforcement Point) Proxy. Orion CB is a context data management system, allowing the operations of registering, updating and querying context data \cite{FIWAREOrion}. It also works with the publish/subscribe communication pattern, allowing the sending of notifications to interested parties when changes occur in the entities of interest. The Keyrock Identity Manager is an identity and access management system, being responsible for registering and querying identity data, providing access tokens to users and validating them in order to authenticate users and allowing access to protected services~\cite{FIWAREKey}. Wilma PEP Proxy is a policy enforcement point responsible for basic authorization, controlling the access to protected services by validating received tokens with the Keyrock Identity Manager~\cite{FIWAREPep}.

For our application, Orion CB receives the energy consumption data from producers (smart meters) and sends notifications to consumers (aggregators); Keyrock and Wilma are used to protect Orion CB and our key manager (Key Vault), doing the authentication and authorization processes.

\subsection{Intel Software Guard Extensions (SGX)}
%------------------------------

Intel SGX is a Trusted Execution Environment (TEE) technology that protects code and data from disclosure or modification~\cite{IntelSGX}. Applications intended to be safe are executed on special protected memory regions such that their code and data are isolated from other software running in the system, even with higher privilege, like operating system (OS).
%, avoiding external communication with the "unprotected world" \cite{SCBR}.
%%TOdo: não faz sentido ficar fazendo referência ao SCBR, parece que não conhecemos os artigos do SGX e estamos baseando nosso entendimento no que outros disseram.
These special regions in the memory are called \textit{enclaves}, which are created and manipulated through a distinct set of processor instructions, with help of a software development kit (SDK) provided by Intel. With these hardware-based capabilities, Intel promises that code and data remain protected even when drivers, OSs or BIOS are compromised.

SGX works with a small attack surface, that is the CPU boundary, preventing direct attacks on executing code or sensitive data in memory. The enclaves work in this boundary, shielding the data and code inside them, and encrypting data when they need to leave the enclave.
%%%(e.g., save data to disk or send them to another application). --> esse exemplo é ruim pois o enclave não faz nada disso automaticamente.
When data return to enclaves, integrity checking is performed. 

Intel SGX also provides a way to enable remote parties to check if an application really executes in a valid enclave of a real Intel SGX processor. This mechanism ensures the authenticity of an enclave, validating, for the remote party, the application enclave's identity. This is possible due to a remote attestation process that is performed following a specific protocol, in which both parts exchange some information in a way the remote application can verify the authenticity of the supposed SGX application by accessing the Intel Attestation Service and checking some information generated in the enclave. At the end of the remote attestation, both sides will have a symmetric shared key that can be used to exchange sensitive information between the remote application and the application running in the attested enclave. The authenticity of the code can also be checked, but the description of this process is out of scope for this paper.

\subsection{Case Study}

This case study is based on a residential environment in which the residences have smart meters to measure their energy consumption for billing purposes. A basic scenario for this application considers the following entities: smart meters, a broker acting as a publish/subscribe system, and some consumers responsible for processing the data.

\begin{figure}[ht!]
  \caption{Basic Scenario for Smart Metering Application.}
  \centering
  \includegraphics[width=6cm, height=5cm]{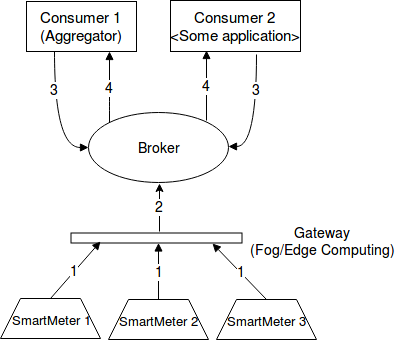}\\
  \label{smartmeteringscenario}
\end{figure}

In this scenario, the smart meters are the data producers and the processing applications are the consumers. A common architecture for this scenario can be seen in the Figure \ref{smartmeteringscenario}. The communication flow between components follows:

\begin{enumerate}

\item The Producers (Smart Meters) send energy consumption data to the IoT Gateway;

\item The IoT Gateway preprocesses these energy consumption data and sends them to a Broker (Publish/Subscribe System);

\item The Consumers register interest on the specific Producers, in the Broker, in order to receive notifications about their energy consumptions;

\item The Broker sends notifications with the Producers' energy consumption data for the Consumers that previously registered interests.
\end{enumerate}

This IoT Gateway can be any device more powerful than the smart meters, for example, capable of processing more complex operations on their generated data or having better communication capabilities. This scenario is often known as fog/edge computing, since the IoT devices are connected to another device with more processing capacity and which is closest to the network edge~\cite{GuanLe17}.

\subsubsection{Aggregation of Smart Meter Data}

For this case study, it was considered that the consumers are data aggregators responsible to process the energy consumption data for billing purposes. This data aggregation is made by summing all the energy consumption data collected for each residence in a specific region. 

To make this aggregation, it is necessary to read all the energy consumption data for all the residences in a region. As stated before, there are approaches to non-invasive estimating what kind of electronic devices or household items are used at specific time and for how long. This technique is called NIALM (Non Intrusive Appliance Load Monitoring) and it is considered a privacy problem since an evil-minded adversary can discover what is done by people living in a specific residence (the time in which people have shower, watch TV or use the microwave, for instance).

\subsubsection{Need for data protection}

Due to the privacy problem mentioned, it is necessary to apply some mechanisms to protect the data generated at producers, avoiding the discovery of the users' behavior in the monitored residences. As the FIWARE does not provide such mechanisms in any security components (Generic Enablers), we had to elaborate a solution to protect the sensitive data using some techniques and our key manager proposed, named Key Vault Generic Enabler, described in the next Section.
\section{Trusted Data Dissemination with FIWARE and Intel SGX}
\label{sec:solution}

Among the current FIWARE Generic Enablers (GEs), there is no specific ones for data protection, enabling users to hide or shield sensitive information and preserving users' privacy. The FIWARE Security modules (GEs) available are useful only for authentication and authorization operations, managing identities and providing access control mechanisms. It was necessary, then, to design and implement a solution that protects the sensitive data 
%of the specific application (energy consumption data aggregation) 
against untrusted or unauthorized third parties, like attackers or even cloud administrators.

\subsection{Threat Model}

Our study application is composed mainly of three parts: the producers (smart meters), the Orion CB (a publish/subscribe system), and the consumers (data aggregators). The attack surface for this application considers the possibility of an attacker gaining access while the data are being transmitted between the parts and the possibility of an attacker gaining complete access to Orion Context Broker and to consumers hosts.

An attacker can intercept the messages sent from the producers (smart meters) to the Orion CB or even the notifications sent from the Orion CB to the consumers (aggregator). Besides, an attacker can get data having access to a consumer host or to our publish/subscribe system (Orion CB). In any case, the sensitive data are compromised. Then, our solution must deal with these threats, avoiding that any attacker in any of these situations can read the sensitive data.

%%% TODO: achei que falta um pouco mais de informação, por exemplo, dava para falar um pouco mais do que o SGX protege e quais as vulnerabilidades dele (por exemplo, side channel attacks, bugs dentro do código). Além disso, é importante lembrar que se alguma informação é usada para roteamento, ela não pode ser criptografada. Se informação sensível for usada para roteamento (ex., estou interessado em medições acima de 100 Watts), então elas não estão protegidas.

Our solution considers that all the producers (smart meters) are secure, thus, we do not deal with the possible existence of fake/crook producers. In any case, it is worth mentioning this possibility could cause only some noise to generated data by valid producers and this would not be a privacy problem for the customers.
% Já está ok acima, não precisa falar de IdM
%Another possible problem that is not considered by our solution is the challenges related to Identity Management (IdM), e.g., privacy, trustworthiness, etc.

\subsection{Principles of the Trusted Solution}

In order to protect the sensitive data, considering the specified threat model, we have used cryptography for data security and privacy along with an identity manager (IdM) with basic access control. Our trusted solution is based on the fact that all the generated data are encrypted in the producers (source), before being transmitted to the publish/subscribe system, and they are only decrypted by a trusted consumer, running on a trusted execution environment (TEE) and with access to the correct key. Furthermore, all the producers and consumers are identified with credentials through the OAuth2 protocol \cite{OAuth}, that is an industry-standard for authorization.

With this architecture, the data encryption in the producers guarantees a secure transmission of the data allowing only a secure processing in a trusted machine, since the consumers must run on a TEE, like SGX. Therefore, we have the data privacy preserved because the sensitive data are decrypted and processed only within a TEE application.

We have designed and implemented a component responsible for generating and storing the keys used by producers and consumers for encryption and decryption: the Key Vault. It also runs on a TEE-enabled machine and is detailed in the next subsection.

In our case study, to validate that the trusted components (Consumers and Key Vault) are running inside SGX enclaves, we perform the remote attestation (RA) process that checks with the Intel if these enclaves are valid and if they are running on a real Intel SGX. This process is done by producers, attesting Key Vault whenever it is necessary to get a public key, and by Key Vault, attesting the consumer whenever it receives a private key request.

%%%TODO: quem faz a atestação? O operador? ou o cliente faz isso automaticamente?

\subsection{The Keyvault Generic Enabler}

%%TODO: producers às vezes é capitalizado, às vezes não. Consumers acho que nunca é.

A Public Key Infrastructure (PKI) was needed to supply appropriated keys to the appropriated entities in a secure way to support our solution. On the one hand, producers must encrypt their data before sending them to Orion Context Broker. On the other hand, consumers must have the appropriate key to decrypt producers' data. Symmetric cryptography is not suitable for this scenario, because an attacker can get access to some producer credentials, be authenticated and request the symmetric key. With this symmetric key, the attacker can decrypt all interested data from the producers.

In order to generate, maintain and distribute keys securely, we developed the Key Vault, which runs in an SGX environment and provides public and private keys accordingly and respectively to producers and consumers from our scenario. The communication with Key Vault is done through Secure Socket Layer (SSL) \cite{SSL}, using HTTPS. The Key Vault provides public keys for authenticated producers and private key for authenticated and attested consumers.

The Key Vault treats the requests from producers and consumers, after being authenticated, and sends the appropriate key (public for producers, private for consumers). As it runs in an SGX enclave, the keys are handled in a secure way. The Key Vault must be attested by producers before sending public keys as well as it must attest the consumers before sending private keys. The private key is only sent to SGX consumers, after the remote attestation properly successful. The private key is encrypted with a shared key that is generated in the remote attestation process. If the the remote attestation of the consumers fails, the private keys are not sent.

\subsection{Trusted Architecture}

Our trusted architecture proposed comprises the following components:

\begin{itemize}

\item Producers - the smart meters, responsible for measuring the energy consumption, with or without some IoT gateway or another device acting as a gateway to process the generated data (Fog/Edge Computing) before sending them to a publish/subscribe system. In the Figure \ref{communication_flow} they are represented as SM, for smart meter, and BBB, for BeagleBone Black acting as a gateway;

\item FIWARE Keyrock IdM - the system responsible for the Identity Management, providing valid OAuth2 tokens for authentication purpose;

%%%TODO: keyrock usa OAuth2 tokens mesmo? No keystone normal dá para usar vários protocolos de autenticação, mas depois de autenticado, o token é de um entre poucos formatos (UUID, PKI, PKIZ, Fernet).
%%% -Sim. Por padrão, ele usa OAuth2. Lembro de ter lido que o desenvolvedor poderia optar por usar um formato de token padrão do Keystone... no entanto, se não estou enganado, o interessado precisaria fazer algumas alterações nem tão triviais. Procurei agora e não achei a referência, mas aqui pode ser visto que tem muita coisa não suportada: https://forge.fiware.org/plugins/mediawiki/wiki/fiware/index.php/Identity_Management_-_KeyRock_-_User_and_Programmers_Guide#Supported_Interfaces

\item FIWARE Wilma PEP Proxy - the Policy Enforcement Point responsible for allowing/disallowing the access to protected services after receiving an OAuth2 token and validates it with Keyrock IdM. This architecture considers two of this component: one to protect Key Vault and another to protect Orion Context Broker;
\item FIWARE Orion CB (Context Broker) - the publish/subscribe system responsible for receiving the producers' data and send them to consumers through notifications;

\item Key Vault - our proposed system responsible for securely managing, storing and distributing the cryptographic keys to producers and consumers;

\item Consumers - the data aggregators responsible to process the energy consumption measures for billing purposes.
\end{itemize}

The communication flow for producers and consumers is detailed next. As mentioned before, the data are encrypted in the producers, which are the smart meters. This is performed as a privacy technique in order to achieve the data privacy. The consumer is responsible for decrypting the sensitive data and processing them in a secure way. All the steps in the communication flow can be seen with a respective number (1 to 5 for producers and 6 to 13 for consumers) in the Figure \ref{communication_flow} and are described bellow:

\begin{figure}
  \setlength\intextsep{0pt}
  \setlength{\textfloatsep}{0pt}
  \setlength{\belowcaptionskip}{0pt}
  \setlength{\abovecaptionskip}{0pt}
  \caption{Producer and Consumer Communication Flow.}
  \centering
  \includegraphics[width=9cm, height=6cm]{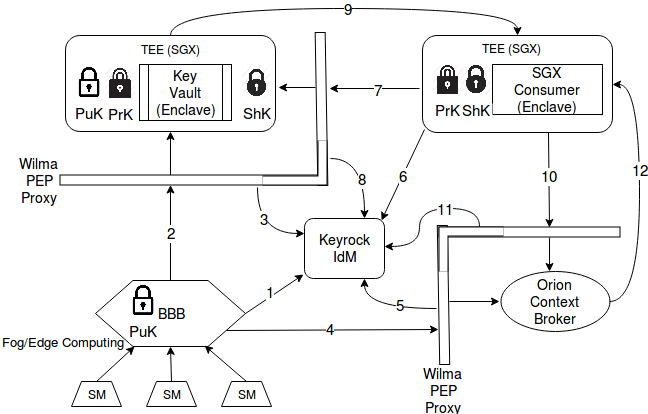}
  \vspace{-8mm}
  \label{communication_flow}
\end{figure}

\begin{enumerate}
\item The producer (smart meter or its gateway) accesses the Keyrock IdM passing valid credentials in order to be authenticated and get an OAuth2 valid token;

\item The producer attests the Key Vault, sends it a request asking for a public key and passing the OAuth2 token obtained with the Keyrock IdM;

\item The Wilma PEP Proxy intercepts the request, checks the validity of the token with Keyrock IdM and forwards the request to the Key Vault if the token is valid. Key Vault then sends back, to the producer, the asked public key (\(PuK\) in the Figure~\ref{communication_flow});

\item With the public key (\(PuK\)), the producer can encrypt the generated data (energy consumption) and send them to the Orion CB;

\item Same as step 3 (Wilma PEP Proxy checks the validity of the token with Keyrock IdM and forwards the request to Orion CB if the token is valid).

\item The consumer, an SGX application, accesses the Keyrock IdM passing valid credentials in order to be authenticated and get an OAuth2 valid token;

\item The consumer sends a request to Key Vault asking for the private key (\(PrK\) in the Figure \ref{communication_flow}) related to the public key sent to producers and passing the OAuth2 token obtained with the Keyrock IdM. Besides the OAuth2 token, the request also informs an endpoint in order to enable the Key Vault to attest that the consumer is a real SGX application (remote attestation);

\item The Wilma PEP Proxy intercepts the request, checks the validity of the token with Keyrock IdM and forwards the request to the Key Vault if the token is valid;

\item Key Vault starts the remote attestation process with the endpoint received and once the consumer is attested, the private key (\(PrK\)) is sent to the consumer. This \(PrK\) is encrypted with the symmetric shared key (\(ShK\) in the Figure~\ref{communication_flow}), before being sent;

\item The consumer sends subscription requests to the Orion CB in order to receive notifications (energy consumptions) from the producers;

\item Wilma PEP Proxy intercepts the request and validates the OAuth2 token with Keyrock IdM. Once the token is valid, the subscriptions are stored in the Orion CB;

\item Consumer receives notifications whenever a producer sends new data (energy consumptions) to Orion CB. As the consumer has the private key (\(PrK\)), it can decrypt data and process them in a secure SGX enclave.

\end{enumerate}

%% TODO: vendo aqui agora, fica claro que o leitor não tem informação suficiente para entender porque rodar no SGX faz diferença. O código pode estar cheio de bugs ou gravar tudo em um arquivo em texto plano. Informações sobre o uso de MRENCLAVE ou MRSIGNER para dizer que aquela versão do código é confiável ou àquele desenvolvedor específico é confiável deveria aparecer lá no início, no modelo de ameaça.

With these steps, the data are transmitted securely from producers to Orion CB. An attacker can not discover what sensitive data are transmitted and stored in Orion since they are encrypted. The messages exchanged with Key Vault are performed through HTTPS.
\section{Evaluation}
\label{sec:eval}

In this section we present the set of experiments we considered to assess the performance of our solution.

\subsection{Experiments Setup}
To evaluate our solution, we deployed all the components of our architecture in two machines equipped with Intel Skylake CPU model i7-6700, 3.4GHz, 8MB cache and 8GB RAM (Dell OptPlex 5040). To have a more realistic scenario, we distributed all the components as follows: one machine with the Producers and the other with Orion CB deployed at a virtual machine, Consumer and Key Vault at separated containers, as well as KeyRock IdM and Wilma PEP Proxy.

After this setup, we have run a set of experiments involving the communication flow explained previously: data generated in the Producers were sent to the Orion CB and it sent the information to the Consumer through notifications. 

\subsection{Scenarios, Metrics, and Parameters}
To analyze the cost of our solution, we considered two scenarios: 
\begin{itemize}
\item Without data security, considering just Producers, Consumer and Orion CB (without any security mechanism);

%%% TODO: cuidado com o termo privacy, como aqui o negócio é binário, eu prefiro chamar de data security. Data security é essencial para privacy, mas é o código do consumidor em si que pode preservar a privacidade ou estragar tudo.

\item With data security, including the Key Vault for key management and distribution, encryption/decryption enabled by these keys, and the authentication/authorization enabled by Keyrock and Wilma PEP Proxy.
\end{itemize}

These scenarios were considered in order to compare our solution with a case using the regular architecture, without employing any security mechanisms. The idea was to verify the cost of the protection added by our proposal.

We have decided to measure latency for both scenarios. This metric was obtained as follows: the elapsed time for some data to leave the Producer and reach the Consumer; as seen before, to accomplish this, the data must leave the Producer and arrive in the Consumer through notifications from Orion CB, in the first scenario; this flow is repeated for the second scenario, considering the key distribution and the encryption/decryption/authentication/authorization processes.

For both scenarios, we have executed 500 publication cycles, to obtain a reliable sample. For each execution, we have marked the time when the data were generated, at the Producer, and time when the data were processed, at the Consumer. Besides the communication flow latency for both scenarios, we also have measured the remote attestation latency, since the remote attestation is an important process performed in the communication with our Intel SGX applications: the Key Vault and the Consumer.

Finally, we also have carried out tests related to the scalability of our solution, in respect to the number of possible producers sending measurements concurrently. We have executed tests with 100, 200, 300 and 400 producers.

\subsection{Results and Discussion}
\begin{figure}
  \vspace{-6mm}
  \setlength\intextsep{0pt}
  \setlength{\textfloatsep}{0pt}
  \setlength{\belowcaptionskip}{0pt}
  \setlength{\abovecaptionskip}{0pt}
  \caption{Latency}
  \centering
  \includegraphics[width=8.5cm, height=6.5cm]{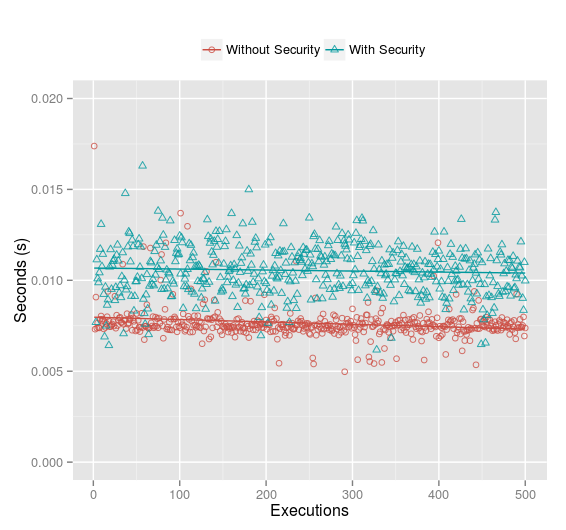}\\
  \label{latency}
\end{figure}

According to achieved results, seen in Fig. \ref{latency}, our proposed solution for data security presents low overhead related to a solution that does not use any security mechanism: approximately 10.5 milliseconds against 7.5 milliseconds on average. The remote attestation process had an average time of 2.5 seconds, but this process is executed only once to get the keys (once producers and consumer have the respective keys, they do not need to communicate with Key Vault).

\begin{figure}
  \setlength\intextsep{0pt}
  \setlength{\textfloatsep}{0pt}
  \setlength{\belowcaptionskip}{0pt}
  \setlength{\abovecaptionskip}{0pt}
  \caption{Latency considering many producers at same time (stress)}
  \centering
  \includegraphics[width=8cm, height=5.5cm]{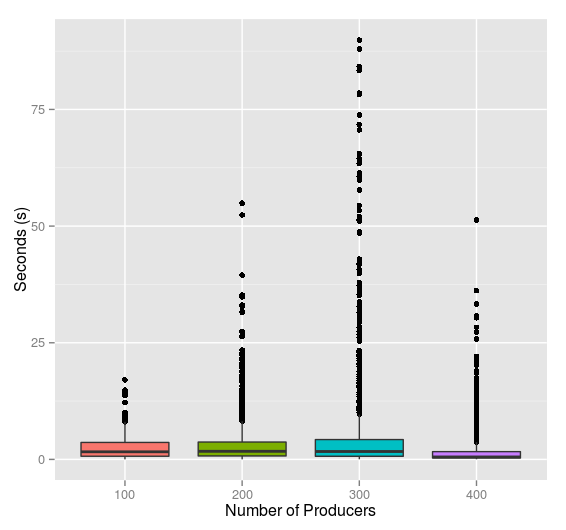}\\
  \label{stresslatency}  
  \vspace{-4mm}
\end{figure}

This solution also presents a good level of scalability, since it has succeed to keep working even with different loads of requests from different numbers of producers at same time: 100, 200, 300 and 400. However, as seen in Fig. \ref{stresslatency}, the higher the number of producers at same time, the higher the latency to process the requests and the lower the number of processed requests.

Considering the threat model applied and experiments, we can say that our solution provides the following features:

\begin{itemize}

\item Integrity -- the produced data are preserved during all the communication flow; they cannot be modified without being detected;

\item Privacy -- the privacy of the customers can be preserved, since the produced data can not be read in any of the communication flow steps, even if the Orion CB is under control of an attacker;

\item Confidentiality -- the confidentiality of the data is preserved because they can not be read/accessed, except by authenticated and authorized producers and consumers;

\item Authentication -- all producers and consumers must use their credentials in order to get access to the protected services and thus successfully perform the communication flow;

\item Secure communication -- data in transit and channel are encrypted.
\end{itemize}

These properties are achieved thanks to: cryptography techniques, TLS/HTTPS protocol and Intel SGX technology. Assuming that the producers credentials are safe/protected, our solution also satisfy the non-repudiation property, in a way any producer can not deny any previous sending of data.

\section{Related Work}
\label{sec:rwork}

In this section we present some works related to security and privacy in IoT, involving general aspects, and Trusted Execution Environments, such as Intel SGX.

Data privacy is always a concern when applications deal with sensitive data. Weber~\cite{Weber2015} and Perera et al.~\cite{Perera-2015} present concerns related to data privacy in IoT, with main challenges and some efforts performed by EU as an attempt to regulate privacy and data protection. For Weber, the main concern is the management of great quantity of generated data in order to secure storage and communication. Skarmeta et al.~\cite{Skarmeta-2014} present requirements that need to be addressed regarding security and privacy challenges in IoT. They propose a distributed capability-based access control that uses public key cryptography and is viable, according to carried out experiments, to deal with complex scenarios in the IoT.

%%TODO: feasible soa estranho para a frase acima, practical?
%% -- Os próprios autores utilizaram esta palavra :)

Werner et al. \cite{Werner17} present a survey on privacy strategies, relating cloud identity management and strategies to achieve good levels of privacy, and listing main features and challenges. The authors proposed a user-centered approach, giving PII (Personally Identifiable Information) control to data owners and encrypting these data to prevent unauthorized access. Henze et al.~\cite{Henze:2016:CAP:2869179.2869387} present an approach that allows users to set their privacy requirements before sending sensitive data to cloud, with an adaptable interface in a transparent way. This is performed during the development process and is also an user-centered approach, with each user owning and operating one or more smart objects (IoT devices).

An end-to-end security architecture for IoT applications is proposed by Vucinic et al~\cite{Vucinic-2015}. They consider a trusted Authorization Server, responsible for providing access secrets to clients (constrained nodes using CoAP protocol) and use encryption and signing mechanisms. Experiments resulted in low energy consumption and latency. Li et al.~\cite{Li2016} propose a  secure communication channel for IoT devices based in a heterogeneous ring signcryption scheme, with identity-based cryptography and a public key infrastructure. Its performance was compared to four other techniques, being more efficient than others and, then, suitable for data transmission in IoT.

Intel SGX is used to implement a Secure Content-Based Routing (SCBR) system \cite{SCBR}, providing privacy preserving to messages, since they are not exposed to unauthorized parties and they are filtered only in a secure enclave. Extensive experiments concluded that SGX adds a limited overhead providing much better performance when compared to other alternatives of Secure CBR. Silva et al.~\cite{Silva2017} presents a solution for privacy and security preservation, also using Intel SGX and considering a smart metering application. Their experiments resulted in a relative low overhead. Guan et al.~\cite{GuanLe17} proposed TrustShadow, a system to shield legacy applications, running on multiprogramming IoT devices, from untrusted OSes. It uses ARM TrustZone technology (TEE) to secure critical applications and it has presented negligible or moderate overhead running real applications.

Sotiriadis et al.~\cite{Sotiriadis} used FIWARE Generic Enablers to implement an architecture that covers the cloud management and the operational features of an application in the health area. This architecture integrates many software modules in order to achieve all the application requirements, including modules for resources and data management, privacy, devices and identity management, cloud resources, etc.

%%%TODO: não temos base para dizer que soluções fim-a-fim não existem, em especial, não discutimos porque AWS e Azure não são completas. Temos um diferencial, SGX, mas nem todo mundo vai achar que isso é o jeito certo (vendor lock-in, disponibilidade limitada). 
%% -- Concordo. Mas o 'there is still a need' remete à necessidade, ainda existente, de soluções assim. De toda forma, alterei o sentido para "there is still space for...", remetendo ao espaço que ainda existe para este tipo de solução.

Security and privacy, as seen in some related works, are still challenges for the IoT systems \cite{wilson2017,Skarmeta-2014,Li2016,Perera-2015}. Although there are good solutions being developed, there is still space for complete solutions end to end, considering all the points we made in our proposal, such as privacy, confidentiality, security, authentication and authorization.
\section{Conclusion}
\label{sec:conclusion}

As highlighted in the related work, data security in the IoT field is still a challenge. In this paper we presented an example that demands such data security: a smart metering system that generates sensitive data. Due to NIALM techniques, this system can be target of privacy attacks, since an attacker can infer behavior from a house by just analyzing its energy consumption measurements.

We then have proposed a solution based on the Key Vault: a new Generic Enabler (GE) responsible for key management and distribution. Key Vault is secure and trusted because it runs on an Intel SGX. In our architecture, Key Vault substitutes the use of a Public Key Infrastructure (PKI) and, thanks to its key distribution and the consequent cryptography used, we achieved data integrity and confidentiality, enabling the development of privacy-friendly solutions.

As seen in the Section \ref{sec:eval}, our solution has a low overhead compared to an architecture that does not consider any data security. Using Key Vault and its capabilities, together with authentication/authorization, our solution had an increase of just 3 milliseconds in the latency of the messages flow, i.e., the elapsed time between producers and consumers considering all the communication flow with the operations of getting keys, encrypting, sending data and decrypting.

%%%TODO: lembrar de completar o XX.

%%% TODO: escalabilidade significa a capacidade de crescer sem limites, embora eu não veja problemas, não testamos.
%Moreover, our solution scales well, since our stress test has reached a number of YY producers sending data almost at same time. We also measured the time spent with the remote attestation process, that is the process performed by any application which wants to validate if a system really runs on a valid/real Intel SGX. The remote attestation process presented a mean time of ZZ milliseconds, what is considered acceptable for our application, since this process is performed basically just one time.

Moreover, our solution does not impose scalability constraints additional to the underlying infrastructure. The Key Vault component and the needed attestation tasks can be easily made scalable due to its simplicity. In addition, our stress test has reached a number of 400 producers sending data almost at same time. We also measured the time spent with the remote attestation (RA) process, that validates if a system really runs on a valid Intel SGX. The RA process presented a mean time of 2.5 seconds, what is considered acceptable for our application, since it is commonly performed just one time for each new producer or consumer, to validate the system or the new member, respectively.

%\subsection{Future Work}
In our solution we used two FIWARE Security GEs: KeyRock Identity Management and Wilma PEP Proxy. Both can be considered as untrusted since they run unprotected in cloud and can be abused by adversaries. As future work, we envisage that these GEs can be implemented with some TEE technology. Furthermore, we can implement Attribute Based Encryption on the Key Vault to control access to keys using intrinsic attributes of the IoT devices. In this case, the access to keys would be controlled by entities' attributes.

\addtolength{\textheight}{-12cm}   % This command serves to balance the column lengths

%\section*{ACKNOWLEDGMENT}

%The preferred spelling of the Put sponsor acknowledgments in the unnumbered footnote on the first page.

%%%%%%%%%%%%%%%%%%%%%%%%%%%%%%%%%%%%%%%%%%%%%%%%%%%%%%%%%%%%%%%%%%%%%%%%%%%%%%%%

\bibliographystyle{IEEEtran}
\bibliography{iscc}

\end{document}